# Natural electromagnetic waveguide structures based on myelin sheath in the neural system

Jiongwei Xue and Shengyong Xu
Department of Electronics, and Key Laboratory for the Physics & Chemistry of Nanodevices, School of Electronics Engineering and Computer Science, Peking University, Beijing 100871, People's Republic of China.
Jiongwei Xue, xuejiongwei@gmail.com
Shengyong Xu, xusy@pku.edu.cn

The *saltatory* propagation of action potentials on myelinated axons is conventionally explained by the mechanism employing "local circuit ionic current flows" between nodes of Ranvier. Under this framework, the myelin sheath with up to 100 layers of membrane only serves as the insulating shell. The speed of action potentials is measured to be as fast as 100 m/s on myelinated axons, but ions move at just 100 nm/s in a 1 V/m electric field. We show here the action potentials, in the form of electromagnetic (EM) pulses, can propagate in natural EM waveguide structures formed by the myelin sheath merged in fluids. The propagation time is mainly cost on the duration for triggering EM pulses at nodes of Ranvier. The result clearly reveals the evolution of axons from the unmyelinated to the myelinated, which has remarkably enhanced the propagation efficiency by increasing the thickness of myelin sheath.

In nature, a high speed for information transmission among different parts of a live creature is one of the crucial factors that determine its survival chance. The information in a biosystem is transmitted either chemically or physically, or by a hybrid combination of both [1]. The velocities for diffusion of molecules in a fluid and for directional movement of ions in an external electric field are both very slow [2]. For instance, in a dilute electrolyte, experimental data show that under an electric field intensity of 1 V/m ions can only move at a speed about 100 nm/s. By sharp contrast, an EM wave propagates in vacuum at a speed of $3\times10^8$ m/s. The transmission of an EM wave can be regarded as transmission of energy, not transport of mass. With limited scattering effects, an EM wave can travel in vacuum or in a uniform dielectric over a long distance without remarkable loss in energy. This is the reason that *Hubble Space Telescope* can receive images of galaxies at the edge of the universe, and a high-quality optical fiber can efficiently transfer data across the *Pacific*. Engineers have also made a variety of transmission lines and waveguides for high-efficiency transmission of EM waves [3,4].

It is well known that the action potential is generated by the transient ion flows passing through ion channels. Maxwell equations show that such transient ion flows emit EM waves in the forms of electrical pulses in the way similar to dipole antennas. And, the electrical pulses can propagate along certain transmission path without employing physical movement of charge carriers. Therefore, it is a wise choice for the nature to transmit information through EM waves.

However, an EM wave is usually strongly scattered in a complex biosystem made of materials with varied phases, densities, dielectric constants, shapes and sizes. The strong scattering processes are sometimes useful for precise imaging [5-7]. The inner part of an axon contains many sub-cellular organelles and cytoskeleton structures such as microtubule, microfilament and neurofilament, thus it has a large resistivity for *dc* current and strongly scatters an EM wave due to its non-uniformity in dielectric constant, shape, size and density of sub-cellular organelles. As a result, the inner part of an axon is not favorable for transmission of electric signals via either *dc* currents or EM waves.

But fortunately, the phospholipid bilayer has a uniform structure. It is the basic framework of the membrane, and such a membrane covers each and every single cell in a biosystem. In this work, we show that by using the membrane as the main frame, EM waveguide structures can be naturally formed in the inner fluidic environment in biosystems, and such EM waveguide structures are efficient for the propagation of action potentials.

**Results**

*The EM waveguide structure formed by electrolytes and dielectrics*

**Figure 1a** schematically shows the way of transmission for a transverse electromagnetic (TEM) wave, in vacuum or a uniform dielectric. Here the electrical field intensity ***E***, the magnetic induction intensity ***B*** and wave vector ***k*** are perpendicular to each other. Along the transmission path, the energy density for a TEM wave, often characterized by $|E|^2$, keeps constant in vacuum and decreases slightly in a waveguide over a long distance.

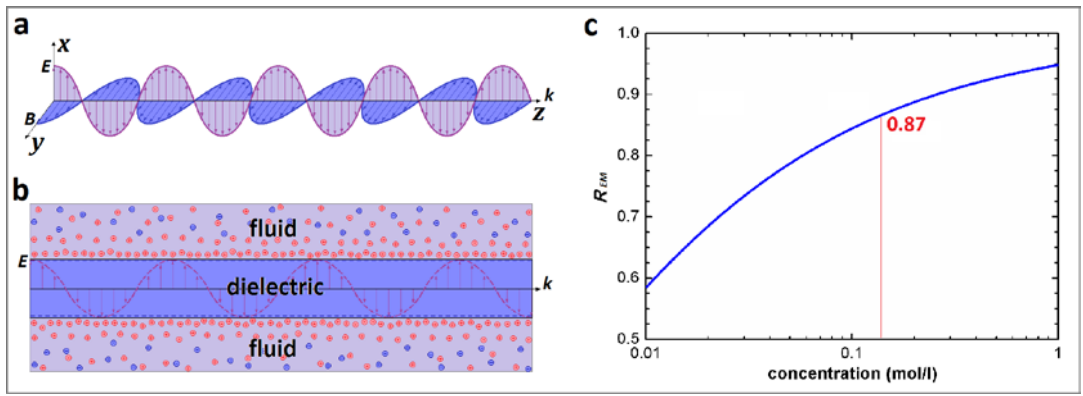

**Figure 1**   a), Schematic diagram of a TEM wave with wave vector of ***k*** travelling in vacuum or a uniform dielectric along ***z*** direction. b), A model of a fluid-dielectric-fluid sandwich EM waveguide structure formed from one layer of dielectric media between two layers of electrolyte. c), The calculated $R_{EM}$ value versus $C_{bulk}$, where the frequency of the EM wave is set at 10 kHz, and the permittivity of the central dielectric layer at 3.0.

It is well known that a conductor-dielectric (or vacuum)-conductor sandwich structure can form a parallel plate waveguide. **Figure 1b** shows the model of an electrolyte-dielectric-electrolyte sandwich EM waveguide structure, which is formed by one layer of dielectric between two layers of electrolyte (see ***Supplementary 1***). The calculated reflection coefficient $R_{EM}$ value versus the concentration of bulk electrolyte $C_{bulk}$ is plotted in **Figure 1c**, where the frequency of the EM wave is arbitrarily set at 10 kHz, and the permittivity of the central dielectric layer at 3.0. In the range of bulk concentration from 0.01 to 1.0 mol/l, $R_{EM}$ increases from 0.58 to 0.95. It suggests that the electrolyte-dielectric-electrolyte sandwich structure works well as an EM waveguide, though not as perfect as a metal waveguide. An EM waveguide can also be formed within a confined fluid channel, as shown in ***Supplementary 2***.

### *The fluid-membrane-fluid sandwich structure as an EM waveguide*

EM waveguide structures can be formed naturally in biosystems with soft materials only. **Figure 2a** shows a typical example, where a membrane, shown with phospholipid bilayer, is sandwiched between two layers of fluid full of ions at both sides. As shown in **Figure 1c**, the reflection coefficient $R_{EM}$ is calculated to be 0.87 at the concentration of 0.15 mol/l, a typically ion concentration in a live biosystem.

Therefore the fluid-membrane-fluid waveguide structure shown in **Figure 2a** can be considered as a parallel plate waveguide. Ideally a parallel plate waveguide does not have dispersion on frequency for propagation of TEM waves. The characteristic impedance of such a waveguide is a constant dependent only on the geometry and material parameters of the structure. For the TEM mode, as showed in **Figure 2b**, in the coordinates where the dielectric layer and the fluid layers are perpendicular to *x* direction and the EM wave is transmitted along *z* direction, the two fluid layers work as the conductor cladding plates, thus $E_y = E_z = 0$, $B_y = 0$, where $E_y$ and $E_z$ are the electrical field intensities along *y* direction and *z* direction, respectively; $B_y$ is the magnetic field intensity along *y* direction. **Figure 2c** schematically shows the field lines of ***E*** and ***B*** in *x-y* plane when viewing along the *z* direction. It gives another reason that why the ion channels are embedded in membrane, generating ion current flows perpendicular to the membrane, for the parallel plate waveguide structure prefer to transmit polarized TEM wave with *E* perpendicular to the membrane.

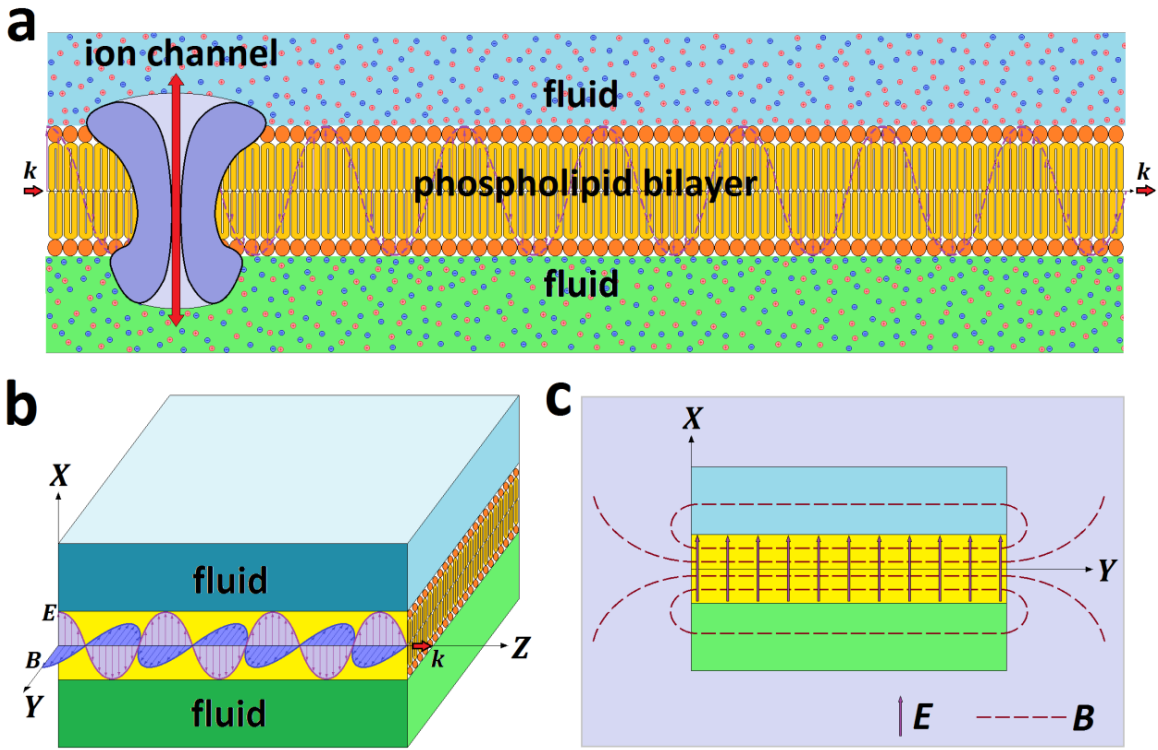

**Figure 2** **a)**, A membrane merged in fluids full of ions at both sides is a typical example of electrolyte-dielectric- electrolyte EM waveguide structure. The dashed purple line indicates a propagating EM wave, but the wavelength of the EM wave is not scaled to the thickness of the membrane. Such an EM wave can be generated by a transient ion current passing through an ion channel embedded in the membrane. **b)**, A schematic three dimensional view of a TEM propagating in the fluid-membrane-fluid waveguide structure along *z* direction, where *E* is perpendicular to the membrane. **c)**, A schematic diagram of the field lines of *E* (represented by arrows) and *B* (represented by dash lines) in *x-y* plane viewing along *z* direction.

*EM coupling between connected membranes*

Next we discuss the propagation of EM waves in the membrane network. In a biosystem when two membranes are close or connected by proteins, a kind of dielectric softmaterial, as shown in **Figure 3a**, an EM wave can inject into the

"X"-shaped structure from one side of any membrane, and comes out from both membranes at the other side. This is an important concept showing that an EM wave can travel through a network of coupled membrane network, as long as the membrane network is surrounded by electrolytes. Therefore, the propagation of dendritic impulses in a neural system is dictated by this principle.

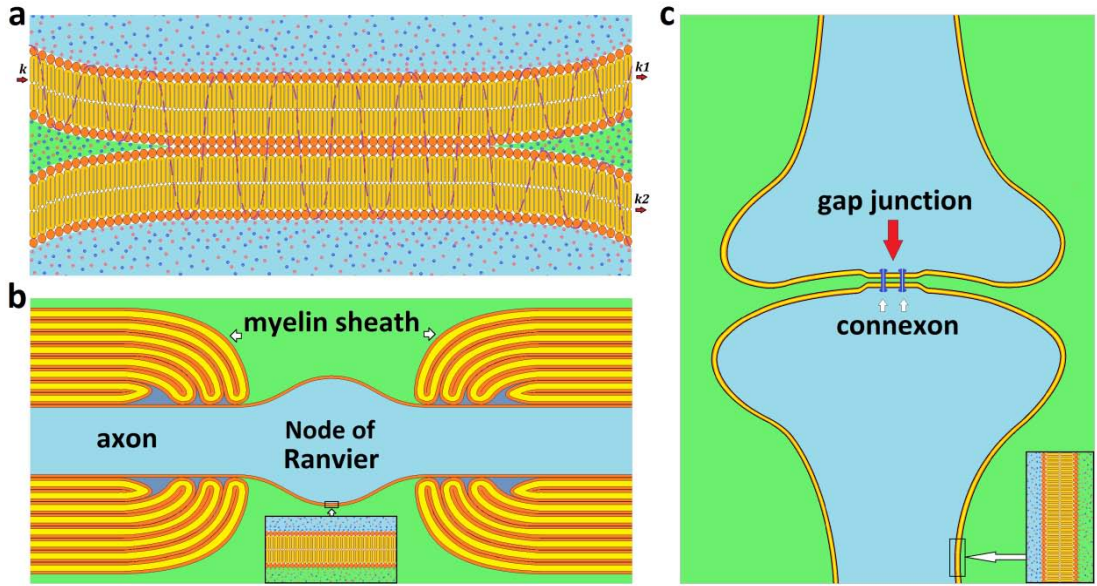

**Figure 3** Cross-sectional images of coupled membrane waveguide structures. **a)**, When two membranes are connected in a biosystem, an EM wave can inject into the "X"-shaped structure from one of the two channels at the left side, and comes out through both channels at the right side, when the membrane network is surrounded with electrolytes. **b)**, Schematic cross-sectional structure of a myelinated axon near a node of Ranvier, where the axon is firmly wrapped with a myelin sheath consisted of multilayered membranes. An EM pulse generated at the node of Ranvier can be transmitted in both the membrane of axon and the myelin sheath. **c)**, An electric synapse, where at the central region there is a gap junction, and the two membranes are connected by connexons. The inset shows the membrane.

**Figure 3b** shows the schematic cross-sectional structure of a myelinated axon near a node of Ranvier, where the axon is firmly wrapped with the myelin sheath, multilayered membranes. An axon is conventionally considered as a transmission line and the myelin sheath acts as the insulating sheath. However it is hard to understand that myelin sheath use even more than one hundred layers of membrane for insulation. Dictated as above, an EM pulse generated at the node of Ranvier can be transmitted both in the membrane of axon and the myelin sheath. Therefore, the myelin sheath acts in fact as the dielectric core of the waveguide structure by adding the thickness of the total membrane layers. **Figure 3c** shows the cross-sectional structure of an electric synapse, where the ends of two opposite axons are coupled with a gap junction. At the gap, with a narrow spacing around 3 nm, the local membranes of two axons are connected by connexons. No ion channels can generate action potentials existing in the gap junction, but experimentally it has been found that action potentials can propagate through the gap junction from one axon to the other without any measurable delay in time.

The mechanism of such a transmission cannot be explained clearly by conventional model with ionic current flows. However, following the model presented in this paper, it is a natural result that the action potential is transmitted in the form of the EM pulse which propagates inside the fluid-membrane-fluid waveguide structure of one axon and passes across the connexon-coupled gap junction and continuously propagates inside the fluid-membrane-fluid waveguide of the other axon. The time delay for such an EM-wave propagation across the coupling junction is in the order of 10 fs, therefore is not measurable.

*Action potentials "saltatory" between nodes of Ranvier on myelinated axons*

Now we can apply the fluid-membrane-fluid waveguide structure to explain the propagation of action potential between neighboring nodes of Ranvier along a myelinated axon, so-called "saltatory" mechanism, which is not clearly explained under conventional framework with local circuit ionic current flows.

**Figure 4** shows schematic pictures of such a process in four frames, where the myelinated axon structure is viewed in its cross-sectional cartoon, and the scales of the axon, axon membrane, myelin sheath and spacing between two nodes of Ranvier do not strictly follow the real scales. The center of the axon is colored in light blue, where three nodes of Ranvier are labeled as A, B, and C. The array of narrow arrows schematically represents the electrical field of the EM pulses transmitted through the membrane of the axon and myelin sheath.

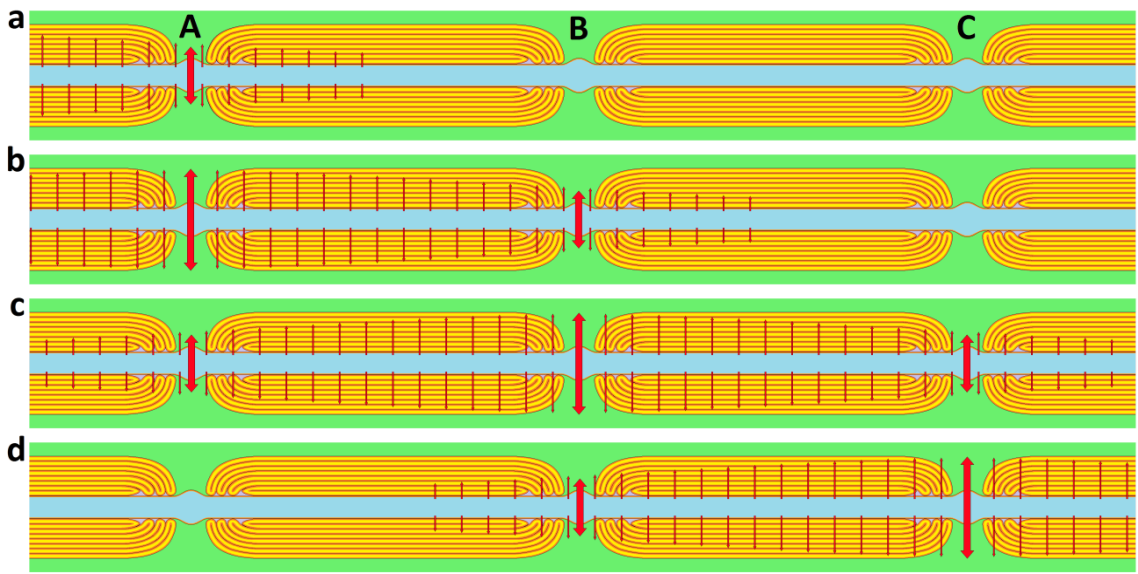

**Figure 4** Schematic cross-sectional cartoon pictures showing the propagation of action potentials along a myelinated axon among three nodes of Ranvier labeled as "A", "B" and "C". **a**), An EM pulse, *E* is represented by the array of narrow arrows, propagates from left side passing through Node A, via the path of fluid-membrane-fluid waveguide structure, causing excitation of ion channels at Node A, as indicated with a thick red arrow. **b**), This excitation causes a transient flow of ions thus generates a fresh EM pulse, and triggers the excitation of ion channels at Node B. **c) & d**), The same process is repeated so that fresh, EM pulses are generated at Node B and Node C, sequentially, and as a result the EM pulse is transmitted from left to right through the membrane of the axon and myelin sheath.

In the first frame, **Figure 4a**, an EM wave propagates from left side passing through Node A, causing excitation of ion channels at this node. The excitation, indicated with a thick red arrow, causes a transient flow of ions perpendicular to the membrane. Thus according to Maxwell Equations, as shown in the second frame **Figure 4b**, it generates a fresh pulse of EM wave, which is similar to the function of a dipole antenna. The power of this EM pulse is determined by the number and acceleration of ions passing through the channel in the transient process. This fresh

EM pulse propagates through the membrane and myelin sheath of the axon, at a speed of ~ $10^8$ m/s, passes through Node B and triggers excitation of ion channels in the membrane at this node, too. Then, as shown in the third frame **Figure 4c**, a fresh EM wave pulse is generated at Node B, which propagates shortly along the axon and triggers excitation at Node C. Again, the fourth frame **Figure 4d** shows the repeated EM wave pulses generated at Node C. If the structure of node of Ranvier, the number and distribution of ion channels, and the triggering mechanism at each ion channel are roughly the same [8-10], we can expect that the fresh EM wave pulses generated at different nodes have the same strength and shape. And this has been proven by many experimental reports.

*The Propagation efficiency for EM waves in different types of axons*

Next we examine the propagation efficiency for EM waves in myelinated and unmyelinated axons of various dimensions.

In a live cell, the skeleton of membrane, the phospholipid bilayer, is usually embedded with certain density of proteins and other molecules, and it consists of intrinsic structural defects. As a result, an EM wave propagating inside the membrane is remarkably scattered, leading to a limited propagation distance. As well as the fluid, full of ions and molecules, is not a conductor as good as the metal.

Now we introduce $L_0$ as the efficient propagation length for an EM wave transmitted along an axon waveguide, defined as in *z* direction. Assume the original electric field intensity $E_0$ attenuates along the propagation path in the way that can be described by $E = E_0 e^{-z/L_0}$. When $z = L_0$, $2L_0$, and $3L_0$, $E \approx 37\% E_0$, 14% $E_0$, and 5% $E_0$, respectively. Therefore, 1-3 $L_0$ is the distance that an EM wave can be transmitted efficiently, beyond which, the energy strength of the propagating EM wave is assumed to be too small to be detected by natural bio-sensors in a biosystem.

We have mentioned that the model shown in **Figure 1b** is similar to that of a parallel plate waveguide. While wrapped on an axon surface, the membrane forms a co-axial geometry, by which the side-leaking effect of EM energy is remarkably eliminated. The unmyelinated axons are usually thicker than myelinated ones. And, experiments show that the thicker the unmyelinated axons, the faster the transmission speed for action potential.

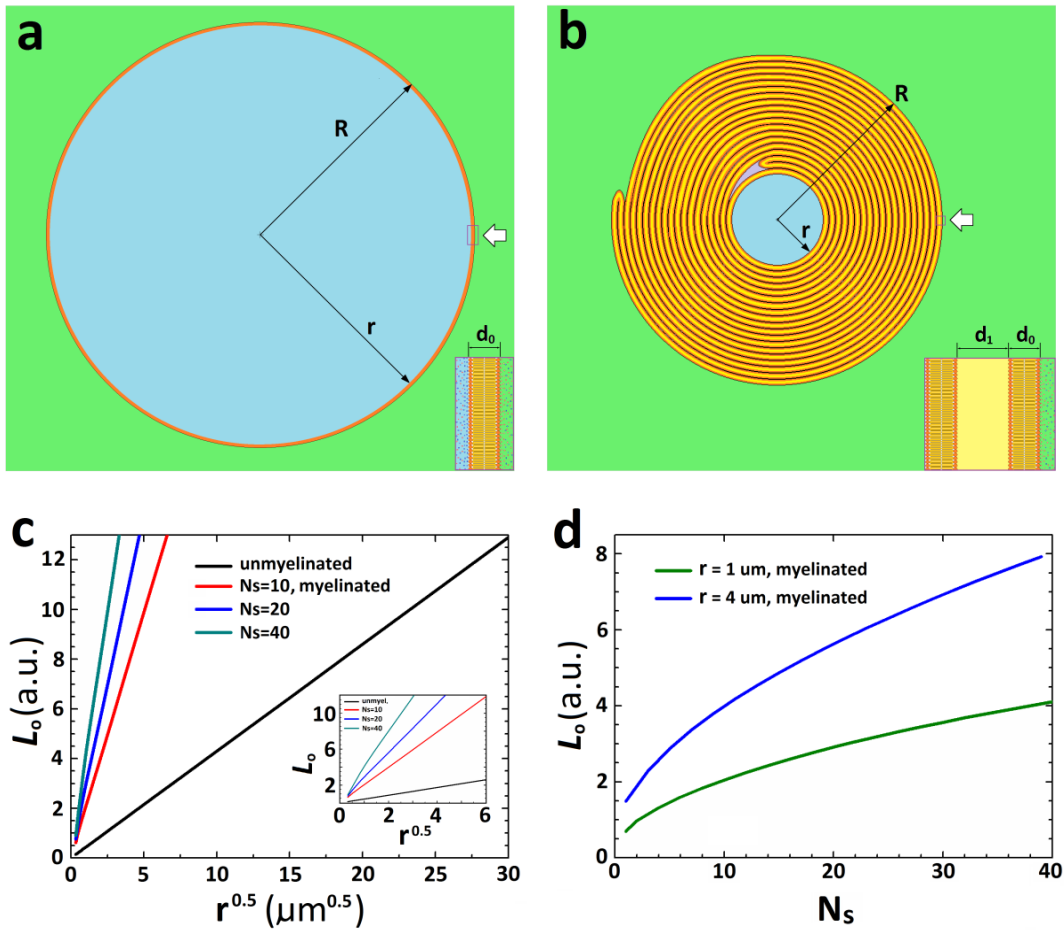

**Figure 5** The dependence of effective propagation length $L_0$ on the axon structure and diameter. **a)** & **b)**, Schematic cross-sections of an unmyelinated axon and a myelinated axon, respectively. Their membrane structure, highlighted with a purple frame in each figure, is a phospholipid bilayer with thickness of $d_0$. The inner and outer diameters are defined as $r$ and $R$. **c)**, Plots of calculated $L_0$ versus $\sqrt{r}$ of an unmyelinated axon (shown in solid black line) and three myelinated axons with $N_s$ of 10 (shown in red line), 20 (shown in blue line) and 40 (shown in green line). **d)**, Plots of calculated $L_0$ versus $N_s$ of two myelinated axons with $r = 1$ $\mu$m (shown in green line) and $r = 4$ $\mu$m (shown in blue line).

**Figure 5a** and **Figure 5b** show schematically the cross-sections of these two kinds of axons, respectively. Both of their membranes are basically made of phospholipid bilayer with thickness of $d_0$ and two protein layers attached to both sides of the phospholipid bilayer. From reported experimental data, $d_0$ is measured to be around 3 nm, and the thickness of the two protein layers, $d_1$, is estimated to be around 7 nm. In the figures the inner and outer diameters of these axons are defined as $r$ and $R$, respectively. Hence $R \approx r+d_0+d_1$ for an unmyelinated axon, and $R \approx r+(2N_s+1)\cdot(d_0+d_1)$ for myelinated axon, where $N_s$ is the number of sheath layers [11].

Based on the waveguide theory, and assuming the protein layer and the phospholipid bilayer have the same dielectric constant, we have calculated the values of $L_0$ versus square root of $r$ for both unmyelinated and myelinated axons. The results are plotted in **Figure 5c** and **Figure 5d**. It qualitatively shows that the transmission speed for action potential linearly increases with $\sqrt{r}$, matching excellently with the well known experimental data that reveal a square root dependence of "length constant" for neural signal transmission on the diameter of axons [12-15]. It also shows, for EM wave transmission a myelinated axon is much more efficient than an unmyelinated axon of the same inner diameter. For instance, the $L_0$ of an unmyelinated axon with $r = 400$ $\mu$m is roughly the same as that of a myelinated axon with 40 layers of sheath, $r = 4.0$ $\mu$m and overall diameter ~ 10 $\mu$m.

**Discussions**

With a geometric structure similar to that of a metallic parallel plate waveguide, the fluid-membrane-fluid waveguide is in favor of propagating TEM waves. In a metallic parallel plate waveguide, the component of a TEM wave with electrical field parallel to $x$ direction can propagate for a long distance, but its component with electrical field parallel to $y$ direction cannot. However, as EM waves propagate in fluids in a way different from that in a metal, so the characteristic nature of a metallic parallel plate waveguide is not fully applicable in the waveguides shown in **Figure 2**. In fluids and membranes, EM waves are expected to show frequency dispersion.

As shown in **Figure 4**, the "*saltatory*" phenomenon observed in myelinated axons is a natural result of propagation of EM pulses between neighboring nodes of Ranvier. It is important to note that, the propagation of an EM pulse along the membrane of axon and the myelin sheath does not require any external bias voltage or directional current of ions along the axon axis either inside or outside the axon. And, this process does not cause the reversal of membrane potential in the internode region. As this region could be as long as 2 mm, indeed the local electric field generated at any node is screened at this distance, therefore the bias for the local circuit ion current proposed in conventional theory does not exist.

The results shown in **Figure 5c** can be considered as a snap shot of such an evolution. In nature the myelinated axons in complex biosystems are developed from unmyelinated axons in simpler biosystems. One layer of membrane for the unmyelinated axon is developed into multilayer of membrane for the myelinated. In the conventional framework of neuroscience, it is hard to explain why the number of membrane layers for the myelin sheath is observed as many as a hundred or even more. However, as shown in **Figure 5c** and **Figure 5d**, in the view of EM waveguide, the reason is clear: The thicker the myelin sheath the better the efficiency of the waveguide. As a result, due to this mechanism, the number of myelinated axons a vertebrate can contain in its spine is a thousand times more than that of unmyelinated ones having the same transmission efficiency, and this is much favorable for advanced communications in the whole body.

Note that in **Figure 5c** and **Figure 5d**, the plotted efficient propagation length $L_0$ is scaled to arbitrary unit (a.u.) due to lack of the exact experimental data. In 1960s, Kanno & Loewenstein conducted experiments on electrical coupling between gland cells. The results showed attenuation of an EM pulse inside a biosystem with a "length constant" (similar to $L_0$) of 0.8-1.2 mm [16]. The well-constructed fluid-membrane-fluid waveguides described in this work is a better path for transmission of EM waves than the random gland cell samples Kanno used in the experiment. Therefore, we can expect an $L_0$ be larger than 1 mm in the best myelinated axons.

The transmission speed, $v_a$, of an action potential signal propagating along an axon is commonly measured to be 1-100 m/s in different kinds of axons. But an EM wave needs only 10 ns to propagate over a 1 m. Therefore almost all the transmission time is allocated to the generation of EM pulses at the ion channels. We define here $\tau$ as the time for the rising edge of a single action potential pulse, which measures 0.1-0.5 ms in various experiments. Hence for myelinated axons when $v_a = 10$ m/s, it leads to an average spacing $\Lambda$ of 1-5mm between two neighboring excitation sources. This rough estimation is well consistent with experimental results showing spacing of 0.2-2 mm between neighboring nodes of Ranvier [17]. And clearly, as the ion channels on one axon are almost identical so that $\tau$ is a constant, in order to increase the exact transmission speed $v_a$, the best strategy for an axon is to reduce the number of nodes of Ranvier per unit length, *i.e.*, to increase the average spacing between two neighboring nodes. As a result, it leads to a simple correlation: $v_a \approx \Lambda/\tau$.

Finally, in the following we discuss a strong supporting evidence for the proposed model that fluid-membrane-fluid EM waveguide structures are the signal transmission paths in a biosystem. In a live human muscle tissue, when the muscle fibers are ordered to work (*e.g.* to shrink), an action potential signal is transferred through an axon to the muscle fibers. At the end of an axon, a synapse is directly contacted to the sarcolemma, an excitable membrane of a muscle fiber. Interestingly, the final sub-cellular structures inside the muscle fiber that reacts with the action potential are a number of myofibrils. Each myofibril is firmly surrounded with a sarcoplasmic reticulum (SR), and each SR is connected via a complex network of T tubules of varied lengths with the sarcolemma, as schematically shown in **Figure 6a** and **Figure 6b**. The action of each myofibril is triggered by voltage-dependent calcium channels, which are special proteins embedded in the SR membrane and can releases a large number of $Ca^{2+}$ ions to the myofibril once triggered.

According to the conventional theories, the action potential can be applied to the sarcolemma via a synapse. However, if this is the case, the transmission of electrical signals inside a muscle fiber faces difficulties: (1), there is no axon or excitable membrane connecting the myofibrils and the sarcolemma; (2), there is no loop for local current flow or pressure gradient, *i.e.*, there is no driving force for the motion of ions or molecules in a T tubule; (3), the number of synapse connected to the muscle fiber is much less than the number of myofibril inside each muscle fiber, so that the length of T tubules from each myofibril to the sarcolemma is different from each other, and (4), it is known that the diffusion speed of molecules in solution, and the motion of ions under an external *dc* field, are both very slow.

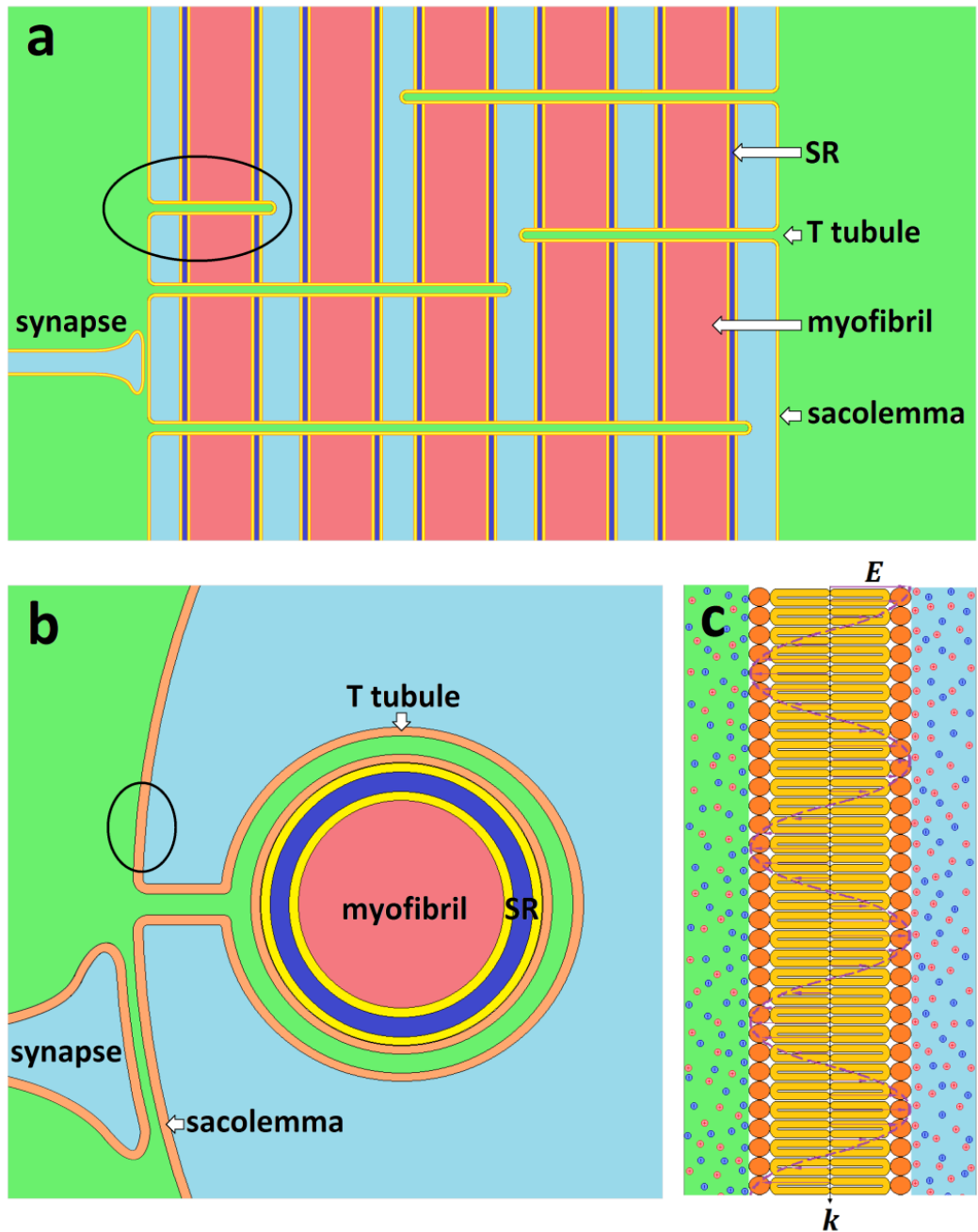

**Figure 6**  **a)**, Schematic diagram of a muscle fiber, where a synapse is connected to the sarcolemma. The myofibrils inside the muscle fiber are connected to the sarcolemma via a complex network of T tubules, and each myofibril is surrounded with a sarcoplasmic reticulum (SR). **b)**, A cross-sectional diagram of the black circle in **a)**, consisting of synapse, sarcolemma, T tubule, SR and myofibril. **c)**, A cross-sectional diagram of the black circle in **b)**, showing an electrolyte fluid-phospholipid bilayer-fluid waveguide structure.

But in a live biosystem a bundle of myofibrils in a single muscle fiber do react together to the same external action potential signal within a time scale of 20-100 ms [18]. This strongly indicates that the same single action potential signal is received by each and every myofibril almost simultaneously. So a consisted explanation is that the signal is not transmitted by molecule diffusion or ion transportation in the T tubules, but by EM waves through the membrane of T tubule. The membrane of T tubule serves as a good EM waveguide, and as a result, although the absolute distance of different myofibrils away from the same synapse-sarcolemma junction varies much, the difference in the transmission times of EM wave is much less than 1 ns, therefore this delay is negligible.

**Conclusion**

Clearly, we have shown that to transmit information in a biosystem by EM waves through the fluid-membrane-fluid waveguide structures is physically feasible and an energy-efficient approach. For transmitting an EM wave *in* a uniform membrane, it needs only a proper source of the EM pulses, such as a transient ionic current flow passing through an ion channel; it does not require any external bias voltage or directional current. The energy dissipation during the propagating of an EM wave is much smaller than that in case of a mass transport.

**Acknowledgements**

We thank W.Q. Sun, J. Tang, C.K. Ong, Bonnie W. Lu, and R.J. Dai for valuable discussions. This work is financially supported by Peking University, the NSFC (Grant No 11074010) and MOST of China (Grant No 2011CB933002).

**Affiliations**

Department of Electronics, and Key Laboratory for the Physics & Chemistry of Nanodevices, School of Electronics Engineering and Computer Science, Peking University, Beijing 100871, People's Republic of China.

**Contributions**

S.Y.X. proposed the concept that an axon works like a fluid EM waveguide and the EM pulse model for the saltatory mechanism on myelinated axons. J.W.X. found the fluid-phospholipid bilayer-fluid waveguide structure and completed the theoretical calculation. J.W.X. and S.Y.X. wrote the manuscript together.

**Competing financial interests**

The authors declare no competing financial interests.

**Corresponding author**

Correspondence to: Shengyong Xu, E-mail: xusy@pku.edu.cn

*Supplementary 1:* **The reflection coefficient $R_{EM}$ of EM waves reflected at the smooth interface of in a uniform dielectric**

In a uniform media, the Maxwell equations for an EM wave can be written in the form of the following:

$$\nabla \times \boldsymbol{E} = -\frac{\partial \boldsymbol{B}}{\partial t} = i\omega\mu\mu_0 \boldsymbol{H} \tag{1}$$

$$\nabla \times \boldsymbol{H} = \frac{\partial \boldsymbol{D}}{\partial t} + \boldsymbol{J} = -i\omega\varepsilon\varepsilon_0 \boldsymbol{E} + \boldsymbol{J} \tag{2}$$

$$\nabla \cdot \boldsymbol{E} = \frac{\rho}{\varepsilon\varepsilon_0} \tag{3}$$

$$\nabla \cdot \boldsymbol{H} = 0 \tag{4}$$

where $\boldsymbol{E}$, $\boldsymbol{B}$, $\boldsymbol{H}$, $\boldsymbol{D}$ and $\boldsymbol{J}$ are vectors for the electric field intensity, magnetic induction intensity, magnetic field intensity, electric displacement and conduction current density; $\rho$, $t$, and $\omega$ are the free charge density, time and circular frequency of the EM wave; $\varepsilon_0$, $\varepsilon$, $\mu_0$ and $\mu$ are the permittivity of vacuum and the media, the permeability of vacuum and the uniform media, respectively. In general, an EM wave propagating in a media with $\rho = 0$ can be written as $\boldsymbol{E}(\boldsymbol{z},t) = \boldsymbol{E}e^{i(\boldsymbol{k}\cdot\boldsymbol{z}-\omega t)} = \boldsymbol{E}e^{-\boldsymbol{\alpha}\cdot\boldsymbol{z}}e^{i(\boldsymbol{\beta}\cdot\boldsymbol{z}-\omega t)}$, where $\boldsymbol{k}$ = $i\boldsymbol{\alpha} + \boldsymbol{\beta}$, with $\alpha$ and $\beta$ being the attenuation constant and phase constant described as:

$$\begin{cases} \beta^2 - \alpha^2 = \omega^2 \mu\mu_0\varepsilon\varepsilon_0 & (5) \\ \boldsymbol{\alpha} \cdot \boldsymbol{\beta} = \frac{1}{2}\omega\mu\mu_0\sigma & (6) \end{cases}$$

For a plane EM wave incident perpendicularly to a conductor,

$$\alpha^2 = \frac{1}{2}\omega^2\mu\mu_0\varepsilon\varepsilon_0 \left(\sqrt{1 + \frac{\sigma^2}{\varepsilon^2\varepsilon_0^2\omega^2}} - 1\right) \tag{7}$$

$$\beta^2 = \frac{1}{2}\omega^2\mu\mu_0\varepsilon\varepsilon_0 \left(\sqrt{1 + \frac{\sigma^2}{\varepsilon^2\varepsilon_0^2\omega^2}} + 1\right) \tag{8}$$

where $\sigma$ is the conductivity of the conductor. In vacuum, $\alpha = 0$.

For an EM wave incident to a surface from material A to material B,

$$\boldsymbol{e_n} \times (\boldsymbol{E_B} - \boldsymbol{E_A}) = 0 \tag{9}$$

$$\boldsymbol{e_n} \times (\boldsymbol{H_B} - \boldsymbol{H_A}) = \boldsymbol{J_{surface}} \tag{10}$$

$$\boldsymbol{e_n} \cdot (\boldsymbol{D_B} - \boldsymbol{D_A}) = \Sigma_{surface} \tag{11}$$

$$\boldsymbol{e_n} \cdot (\boldsymbol{B_B} - \boldsymbol{B_A}) = 0 \tag{12}$$

where $\boldsymbol{E_A}$, $\boldsymbol{H_A}$, $\boldsymbol{D_A}$ and $\boldsymbol{B_A}$, are the electrical field intensity, magnetic field intensity, electric displacement and magnetic induction intensity in material A, respectively, and $\boldsymbol{E_B}$, $\boldsymbol{H_B}$, $\boldsymbol{D_B}$ and $\boldsymbol{B_B}$ are the parameters for material B. $\boldsymbol{J}_{surface}$ is the surface free current density, $\Sigma_{surface}$ the surface free charge density, and $\boldsymbol{e_n}$ the normal vector pointing to the inside of material B.

Defined $\boldsymbol{E}$, $\boldsymbol{E'}$ and $\boldsymbol{E''}$ are the electrical field intensities for the incident, reflected and transmitted EM waves, respectively. So $\boldsymbol{E_A} = \boldsymbol{E}+\boldsymbol{E'}$, and $\boldsymbol{E_B} = \boldsymbol{E''}$. The reflection coefficient $R_{EM}$ is defined as:

$$R_{EM} = \left|\frac{\boldsymbol{E'}}{\boldsymbol{E}}\right|^2 \tag{13}$$

An ideal waveguide has $R_{EM}$ of 1, while for practical waveguides $R_{EM} < 1.0$, *i.e.*, part of the EM wave energy is lost in the conductor by Joule heating, so that the strength of the EM wave decreases along the transmission path. And, the EM wave energy is also dissipated via scattering process in the dielectric.

It has been proven that an EM wave can only penetrate a thin layer of metal. The thickness of this thin layer is defined as the penetration depth $\delta$, and for a good conductor $\delta \approx \sqrt{\frac{2}{\omega\mu\mu_0\sigma}}$. As a result, an incident EM wave completely reflects at the surface of a perfect conductor, similar to an incident light reflected by a mirror. Thus an EM wave can be trapped between two walls of conductors. A conductor-dielectric (or vacuum)-conductor sandwich structure can form a waveguide, so that an EM wave can travel inside the dielectric layer with little energy leakage. The efficiency of a waveguide for confining the energy of an EM wave in its inner space can be characterized by $R_{EM}$. An ideal waveguide has $R_{EM}$ of 1, while for practical waveguides $R_{EM} < 1.0$, *i.e.*, part of the EM wave energy is lost in the conductor by Joule heating, so that the strength of the EM wave decreases along the transmission path. And, the EM wave energy is also dissipated via scattering process in the dielectric media. Along the transmission path, the energy density for a plane EM wave, often characterized by $|\boldsymbol{E}|^2$, keeps constant in vacuum and decreases slightly in a waveguide over a long distance.

According to the Nernst-Planck Equation, the conduction current density $J$ of the electrolyte fluid layer along $z$ direction can be written as:

$$J(z) = \sum_i N_A e_0 Z_i \left[ -D_i \frac{\partial C_i(z)}{\partial z} - \frac{F}{RT} D C_i \frac{\partial \varphi}{\partial z} + C_i V(z) \right] \tag{14}$$

where $N_A$ is Avogadro's constant, $e_0$ being the proton charge; $D_i$, $Z_i$, and $C_i$ are parameters for the $i^{th}$ kind of ions, namely the coefficient of diffusion, valence and concentration, $V(z)$ is the fluid velocity along $z$ direction; $F$ being Faraday constant, $R$ the gas constant, $\varphi$ the electrical potential, and $T$ the absolute temperature. We consider an electrolyte solution in which the concentrations of cations and anions are the same, defined as $C_{bulk}$. If there is no gradient of concentration in $z$ direction and $V(z) = 0$, in a uniform applied electric field $E$, the $dc$ conductivity of bulk electrolyte, $\sigma_{bulk}(0)$, is estimated to be $\sigma_{bulk}(0) = 2\mu_M N_A e_0 C_{bulk}$, where $\mu_M$ is the mobility of ions [1,2].

The $ac$ conductivity $\sigma(\omega)$ of a bulk electrolyte fluid can be described as [3]:

$$\sigma_{ac}(\omega) = \sigma_1(\omega) + i\sigma_2(\omega) \tag{15}$$

$$\sigma_1(\omega) = \frac{\sigma(0)}{1 + (\omega Dm/k_B T)^2 - [2\omega^2 - \omega^4/\Omega^2]/\Omega^2} \tag{16}$$

$$\sigma_2(\omega) = \frac{\omega \sigma(0)[Dm/k_B T - k_B T(\Omega^2 - \omega^2)/Dm(\Omega^2)^2]}{1 + (\omega Dm/k_B T)^2 - [2\omega^2 - \omega^4/\Omega^2]/\Omega^2} \tag{17}$$

where $\Omega$ is the Einstein Frequency, $k_B$ being Boltzmann's constant. Only the conduction current, which has the same phase with that of the electric field, consumes the energy.

*Supplementary 2:* **An EM waveguide structure consisting of two heavily charged dielectric layers and a fluid with a large gradient of ion concentration**

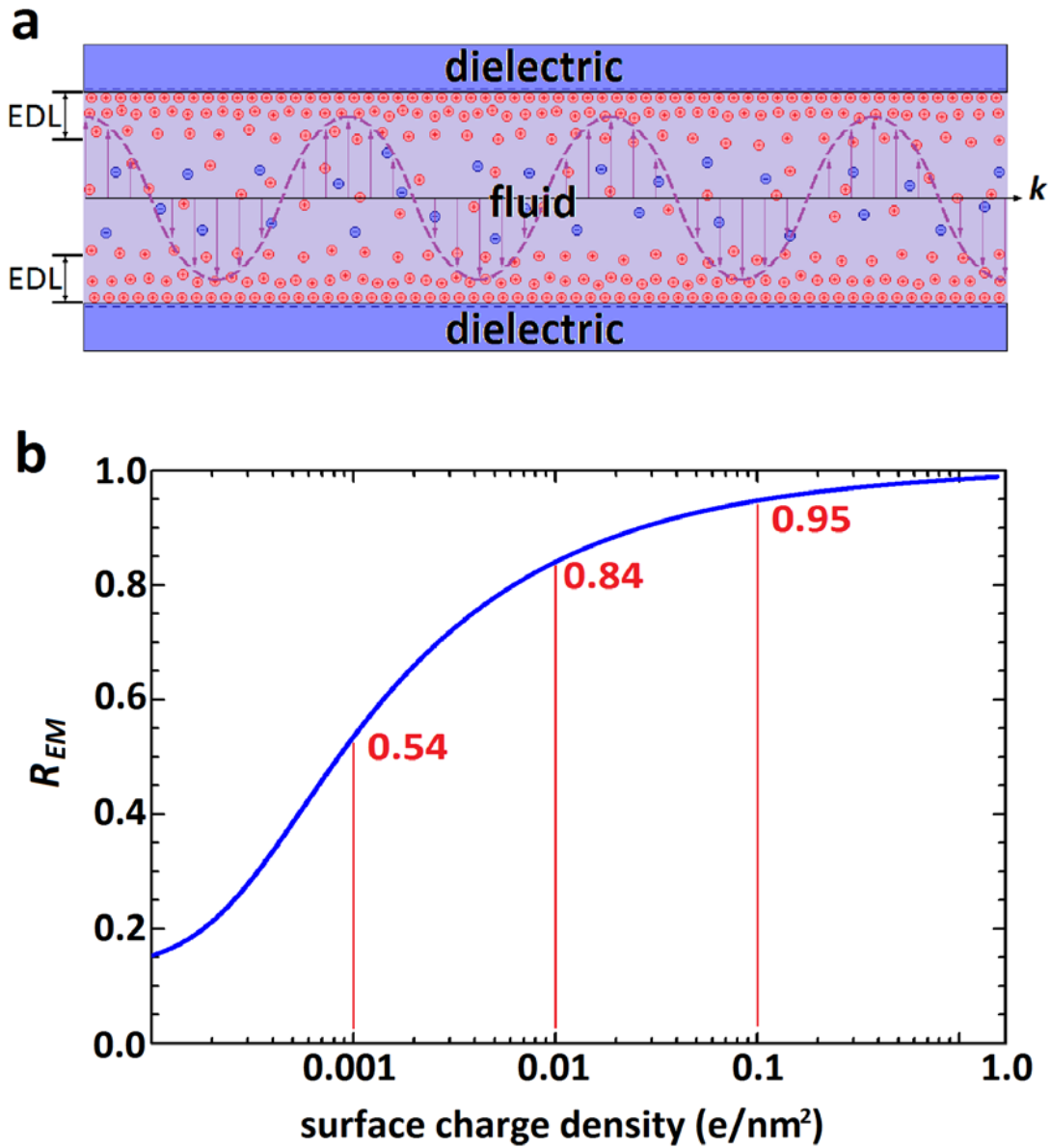

**Figure *S1*** **a)**, Schematic model of another possible fluid EM waveguide, where a fluid layer is trapped between two dielectric layers, and a large gradient of ion concentration exists from the edge of the fluid towards the center. **b),** A plot of calculated $R_{EM}$ value versus surface charge density $\Sigma_s$ for the waveguide model proposed in **a)**, where $C_{bulk}$ is set at $1.0 \times 10^{-7}$ mol/l.

Briefly, in a practical system, when a solid surface is in contact with electrolyte, it is often charged with a surface charge density of $\Sigma_s$. In a dilute electrolyte solution, the average concentration of counterions in the electrical double layer (EDL) [4] is much larger than that in the bulk solution, so a large gradient of ion concentration exists from the edge of the channel towards the center. The thickness of an EDL can be estimated to be the Debye length, $1/\kappa$, where:

$$\kappa^2 = \frac{2e_0^2 N_A C_{bulk}}{\varepsilon_0 \varepsilon k_B T} \tag{18}.$$

As a result, the EDL has a much higher conductivity when compared to the dilute bulk solution, and it can serve as the conducting cladding layer of a waveguide. The conductivity of EDL, $\sigma_{EDL}$ can be revealed from $\Sigma_s$ or the zeta-potential, $\zeta$ [5]. **Figure *S1b*** shows that, this kind of fluidic waveguide only works well under the condition when the surface charge density is high and the ion concentration of bulk solution is low. By contrast, our calculation shows that when $C_{bulk}$ is bigger than 0.1mol/l, the EDL phenomenon has negligible effect on $R_{EM}$ for the structure shown in **Figure 1b**.

In this waveguide structure, a large gradient of ion concentration exists from the edge of the channel towards the center. In practical systems, when a solid surface is in contact with electrolyte, it is often charged with a surface charge density of $\Sigma_s$. In a dilute electrolyte solution, the average concentration of counterions in the EDL is much larger than that in the bulk solution, as shown in **Figure *S1a***, where the channel surface facing the electrolyte is assumed negatively charged. The thickness of an EDL can be estimated to be the Debye length, $1/\kappa$. Compared to the dilute bulk solution, the EDL has a much higher conductivity, and it can serve as the conducting cladding layer of a waveguide. Thus the fluidic channel structure shown in **Figure 1c** is indeed an EM waveguide. The conductivity of EDL, $\sigma_{EDL}$ can be revealed from the surface charge density $\Sigma_s$ or the zeta-potential, $\zeta$. When the surface charge density is high and the ion concentration of bulk solution is low, the waveguide works well. **Figure *S1b*** shows an extreme example, where $C_{bulk}$ is set at $1.0\times10^{-7}$ mol/l and $R_{EM}$ is plotted versus $\Sigma_s$. In this case, one sees that $R_{EM}$ increases with $\Sigma_s$, from 0.54 at 0.01 $e$/nm$^2$ to 0.95 at 0.1 $e$/nm$^2$.